\documentclass[12pt]{iopart}

\usepackage{iopams}  
\usepackage{color}
\usepackage{graphicx}
\usepackage{dcolumn}
\usepackage{bm}
\usepackage{epsfig}
\usepackage{amssymb}

\usepackage{color}

\begin{document}

\title{Critical points of the random cluster model with Newman-Ziff sampling}
\author{Tolson H. Bell$^{1,2}$, Jerrell M. Cockerham$^{1,3}$, Clayton M. Mizgerd$^{1,4}$, Melita F. Wiles$^{1,5}$ and Christian R. Scullard$^6$}
\address{$^1$Institute for Pure and Applied Mathematics, UCLA, Los Angeles, CA 90095, USA}
\address{$^2$Department of Mathematical Sciences, Carnegie Mellon University, Pittsburgh, PA 15213, USA}
\address{$^3$Mathematics Department, Rice University, Houston, TX 77005, USA}
\address{$^4$Department of Mathematics, Williams College, Williamstown, MA 01267, USA}
\address{$^5$Department of Mathematics, The College of Wooster, Wooster, OH 44691, USA}
\address{$^6$Lawrence Livermore National Laboratory, Livermore, CA 94550, USA}

\eads{\mailto{thbell@cmu.edu}, \mailto{jc184@rice.edu}, \mailto{cmm12@williams.edu}, \mailto{mwiles22@wooster.edu}, \mailto{scullard1@llnl.gov}}

\date{\today}

\begin{abstract}
We present a method for computing transition points of the random cluster model using a generalization of the Newman-Ziff algorithm, a celebrated technique in numerical percolation, to the random cluster model. The new method is straightforward to implement and works for real cluster weight $q>0$. Furthermore, results for an arbitrary number of values of $q$ can be found at once within a single simulation. Because the algorithm used to sweep through bond configurations is identical to that of Newman and Ziff, which was conceived for percolation, the method loses accuracy for large lattices when $q>1$. However, by sampling the critical polynomial, accurate estimates of critical points in two dimensions can be found using relatively small lattice sizes, which we demonstrate here by computing critical points for non-integer values of $q$ on the square lattice, to compare with the exact solution, and on the unsolved non-planar square matching lattice. The latter results would be much more difficult to obtain using other techniques.
\end{abstract}

\maketitle

\section{Introduction}
The random cluster model is of fundamental importance in statistical mechanics, with its special cases including percolation and the Ising and Potts models \cite{Potts52,FK1972,Wu82}, and touches a wide range of subject areas. Percolation, for example, has proven to be fertile ground for many fascinating rigorous \cite{Kesten82,Smirnov,Schramm2001} and non-rigorous \cite{Cardy92} results in mathematics, while also serving as a simple model of forest fires \cite{vonNiessen1986,Loehle1996} and epidemics \cite{Ziff2021}. The Potts model, aside from being a general model of ferromagnetism, likewise has found a wide range of surprising applications. For only one example, the three-dimensional 3-state model is used to study the heavy quark limit in lattice quantum chromodynamics \cite{Svetitsky1982,Alford2001}. Thus, the search for efficient numerical methods for the random cluster model is important, and new techniques can potentially impact many different fields.

Critical slowing down has long troubled Monte Carlo simulations of the Potts model. The Swendsen-Wang (SW) algorithm \cite{SwendsenWang}, which greatly mitigates this problem, is now the basis for most modern Potts model simulation techniques \cite{Wolff1989,Machta,Ferrenberg1988}. However, it is restricted to integer $q$ and is therefore not completely general, as there are many applications for which non-integer $q$ is interesting. For example, the range $0 \le q \le 1$ was shown to be in the same universality class as the gelation transition in branched polymers \cite{Lubensky1978}, $q=1/2$ is related to the dilute spin glass problem \cite{Wu82}, and in three dimensions the transition from a second- to first-order critical point occurs for $2 <q <3$ \cite{Grollau2001}. In addition, conformal field theory \cite{DiFrancesco} makes many predictions conjectured to be exact in the continuous range of $q$. While methods exist that handle real $q$ \cite{Sweeny1983,Elci2013,Chayes1998,Gliozzi2002}, there are often trade-offs involved either in efficiency or ease of use compared to SW. 

Here, we present a Monte Carlo scheme, based not on any of these previous techniques but on the Newman-Ziff (NZ) approach to percolation. Advantages of our method are that a) it works for real $q>0$ and that b) results for an arbitrary number of different $q$ can be obtained from a single simulation. One price is that the memory requirement is somewhat steeper than other methods, scaling as $\mathrm{O}(L^3)$ for a lattice with $L$ vertices on an edge. Another, more serious, disadvantage is that as $L$ is increased, configurations that are important when $q>1$ are increasingly poorly sampled. So, although this method does not exactly have the critical slowdown problem, we are nevertheless limited in the size of the systems we can study. On the other hand, because the critical polynomial \cite{Scullard2012,ScullardJacobsen2012,JacobsenScullard2013} provides very accurate estimates of transition points, even on relatively small graphs, it can be used very effectively with the present Monte Carlo scheme. In the standard polynomial method, the polynomial, or more often just its root \cite{Jacobsen15}, is computed via an analytic technique \cite{Jacobsen2014} and, for two-dimensional planar lattices, provides accuracy far surpassing what is currently possible with Monte Carlo methods \cite{ScullardJacobsen2020}. However, relying as it does on a transfer matrix calculation, it is somewhat challenging to generalize to non-planar, not to mention higher-dimensional, lattices. Additionally, even in the planar case, it is rather more complex to implement than a typical Monte Carlo algorithm and one may prefer to trade some accuracy for simplicity.
 
\section{Algorithm}
In a configuration of the random cluster model, each edge of a lattice graph is chosen to be open or closed. The partition function with edge weight $p$ is given by \cite{FK1972}
\begin{equation}
 Z=\sum_{\{\omega\}} p^n (1-p)^{N-n} q^{C(\omega)}
\end{equation}
where $n$ is the number of edges present in the configuration $\omega$, $C(\omega)$ is the number of connected components (including isolated vertices) in $\omega$, and the sum is over all configurations. It is convenient to write $Z$ in the form
\begin{equation}
 Z=\sum_{n=0}^N \sum_{C=1}^{L^2} p^n (1-p)^{N-n} q^C \Gamma_n(C), \label{eq:Z}
\end{equation}
for a two-dimensional lattice with $L^2$ vertices and $N$ edges, and where $\Gamma_n(C)$ is the number of configurations consisting of $n$ edges and $C$ connected components. This function is obviously not known analytically in general, but it is a simple matter to estimate it during the simulation. The probability of some event $A$ is then
\begin{equation}
 P(A)=\frac{1}{Z} \sum_{n=0}^N \sum_{C=1}^{L^2} p^n (1-p)^{N-n} q^C \Gamma_n(C) P(A|n,C) \label{eq:PA}
\end{equation}
where $P(A|n,C)$ is the probability of $A$ for fixed $n$ and $C$. Therefore, to compute the probability of an event $A$, the quantities $P(A|n,C)$ and $\Gamma_n(C)$ can be estimated for all $n$ and $C$ and the results combined in Equation (\ref{eq:PA}). All configurations with a given $n$ and $C$ have the same probability, and thus they might be said to constitute a microcanonical ensemble. More concretely, we have
\begin{equation}
 P(A|n,C)=\frac{\Omega_n(A,C)}{\Gamma_n(C)}
\end{equation}
where $\Omega_n(A,C)$ is the number of configurations consisting of $n$ edges and containing $C$ clusters for which the event $A$ occurs. Random sampling is therefore used only to count configurations. The numerical approach encapsulated in Equation (\ref{eq:PA}) is the generalization to arbitrary $q$ of that of Newman and Ziff \cite{NewmanZiff2000,NewmanZiff} (NZ), to which this method reduces when $q=1$. Their sampling procedure will be used here as it is straightforward to implement and hardly needs any generalization for the present purposes. Although we describe it below, we recommend the reader consult their paper for more details.

The sampling algorithm for a single run works as follows. First, the lattice is empty, with all edges closed. We then add edges one by one in a completely random order until they are all open. At step $n$, we find the number of clusters, $C$, which is simple to do; adding an edge can only reduce $C$ by one or leave it unchanged, and when $n=0$ we have $C=L^2$. Performing multiple runs we then find $f_n(A,C)$, the fraction of runs for which the event $A$ occurs at a given $n$ and $C$. The estimate of the number of configurations is then
\begin{equation}
 \Omega_n(A,C)={N \choose n} f_n(A,C) .
\end{equation}
The only difference between runs is the order in which the edges are added to the system. Permutations of the edge order are handled in exactly the same way as in NZ, and in fact they provide a function written in C called \verb!permutation()! which will work unchanged here. Adding all $N$ edges while tracking clusters takes a time that is essentially $\mathrm{O}(N)$ \cite{NewmanZiff}. Although we do not have independence between step $n$ and step $n+1$ within a single run, all that matters is that there be independence between runs and that all configurations with a fixed $n$ and $C$ should appear with equal probability, which is plainly the case here. We can now see how the critical slowdown and the restriction to integer $q$ are evaded by this algorithm. The configuration sampling knows nothing about the temperature (the probability $p$ in our case) or $q$, but rather these appear as parameters in a convolution, Eq. (\ref{eq:PA}), that occurs after the sampling has been done. For the same reason, we get results for every value of $q$ in a single simulation.

To identify and merge clusters we use the standard \cite{Galler1964,Sedgewick} tree-based union/find described in section IIB of NZ. Although we define configurations by their edge states, clusters are tracked only on the vertices. The key idea is that every cluster has a unique root site and that any given site can be quickly traced to its root. To achieve this, we define an array of integer pointers, \verb!ptr[i]!, which for vertex $i$ either gives its parent site in the cluster or, if $i$ is a root, its value is the negative of the size of the cluster. Thus, every site is connected by a path of parents to the root. So if an edge joins the two sites $i$ and $j$, we take the following steps
\begin{enumerate}
 \item Find the roots of $i$ and $j$ by traversing their trees.
 \item If $i$ and $j$ have the same root, they are in the same cluster. This situation will be important below, but as far as the cluster configuration goes we need do no more. 
 \item If $i$ and $j$ are different, determine which cluster is smaller. If, say, $j$ is in the smaller cluster, we make the root of $j$ point to the root of $i$ and add the size of the smaller cluster to the larger one by adjusting \verb!ptr[i]!. Obviously, if both clusters are the same size we can choose arbitrarily which cluster to add to which.
\end{enumerate}
The algorithm is most efficient if all sites point to their roots. Thus, after we have found the roots with step 1, it is worthwhile to backtrack over the path and set all pointers to point directly to the root we have just found. See NZ for a fuller discussion of this path compression.

\section{Critical polynomials}
\begin{figure}
\begin{center}
\includegraphics{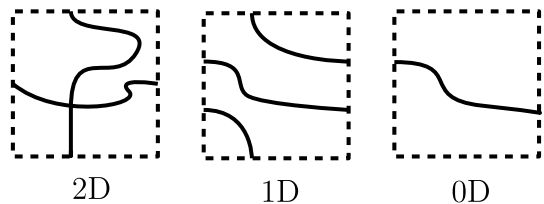}
\caption{The three possible wrapping events on a periodic lattice, or basis.}
\label{fig:wrapping}
\end{center}
\end{figure}

To locate the critical point, we estimate the critical polynomial. This is defined (up to a sign) on a doubly periodic finite lattice, or basis, $B$ by \cite{JacobsenScullard2013}
\begin{equation}
 P_B(p,q) \equiv P(\mathrm{2D};p,q)-q P(\mathrm{0D};p,q) \label{eq:PB}
\end{equation}
where $P(\mathrm{2D})$ is the probability that there is a cluster that wraps both dimensions and $P(\mathrm{0D})$ is the probability that no cluster wraps either. A third possibility, which we will encounter later, is referred to as 1D and is the event that a wrapping spans only one dimension. See Figure \ref{fig:wrapping} for examples of these events. That the root in $[0,1]$ of $P_B(p)$ provides an estimate for the transition point is well established \cite{ScullardZiff08,ScullardJacobsen2012,Mertens2016} and is a consequence of universality. 

The advantage of using the critical polynomial, and the reason for its natural compatibility with our Monte Carlo scheme, is that it can be used with small lattices. For lattices for which the critical points can be determined analytically using a duality argument, such as the square lattice, equation (\ref{eq:PB}) will give the exact answer for any $L$. For an unsolved problem, the critical polynomial does not give the exact answer for any size of basis. However, the estimates it provides are generally very accurate even when the basis is of a size normally considered small for Monte Carlo. For example, for the kagome lattice, a basis of size $L=4$ already produces estimates of the percolation threshold accurate to six digits \cite{ScullardJacobsen2012}. By the time $L \sim 20$ a polynomial prediction for a given system (provided it does not possess some long-range non-planar features) is generally accurate enough that it cannot be checked with Monte Carlo. Thus, using these relatively small lattices is perfectly acceptable in the present approach.

Estimates are obtained by computing averages of $P(\mathrm{2D}|n,C)$, $P(\mathrm{0D}|n,C)$ and $\Gamma_n(C)$ and combining them with Eq. (\ref{eq:PA}). As a function of $p$, $P_B(p,q)$ is monotonic in $p$, taking the value $-q$ at $p=0$ and 1 at $p=1$. A simple root-finding algorithm suffices to find the estimate for $p_c$, and here we use the bisection method. By making good initial guesses, perhaps derived from smaller-sized lattices, one can minimize the number of evaluations needed by this root-finding step. 

\begin{figure}
\begin{center}
\includegraphics{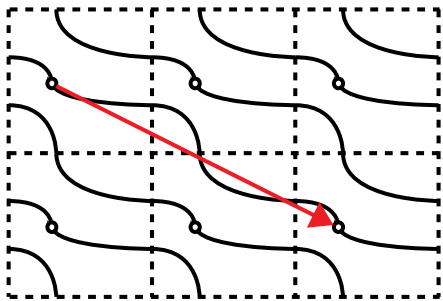}
\caption{The 1D wrapping event of Figure \ref{fig:wrapping} tiled to demonstrate the meaning of the connection vector, indicated by the red arrow. The circle denotes the root of the cluster. The connection vector is $(2,-1)$ in this case.}
\label{fig:convec}
\end{center}
\end{figure}

The next piece of the algorithm is the identification of the different wrapping events. This is done by using a variation of an approach first suggested by Machta et al. \cite{Machta}. For each site $i$, in addition to the pointers \verb!ptr[i]!, we also associate \verb!x[i]! and \verb!y[i]!, the $x$ and $y$ lattice displacements to the root site of the vertex $i$. Upon joining two vertices that are already in the same cluster, if at least one vertex' displacement vector points through a boundary then this operation might produce a wrapping cluster. That is, the root of the cluster can now be connected to itself via a path that winds through one or both of the periodic directions. The exact manner of this winding can be encoded in a connection vector, as shown in Figure \ref{fig:convec}. It is a simple matter to compute this vector for each join we perform. If we are joining vertices, say 1 and 2, for which the vector connecting 1 to 2 is ${\bf u}_{12}$, and their root displacement vectors are ${\bf d}_1$ and ${\bf d}_2$ then the connection vector is given by
\begin{equation}
 {\bf c}={\bf d}_2-{\bf d}_1+{\bf u}_{12}  \mathrm{\ mod\ } L
\end{equation}
(note that the overall sign is irrelevant). If the connection vector is zero, then we have created no new wrapping. Upon forming a 1D wrapping, we must store the connection vector associated with that cluster \footnote{As a practical matter it can be stored in the same array as the displacement vectors, as we know the displacement vector for the root is zero and its entry can be used for something else}. Call this vector ${\bf c}_0$, and if upon later joining two further vertices of this cluster we form a new connection vector, ${\bf c}_1$, then we must determine the span of the two vectors. One way is to evaluate the cross product, and if
\begin{equation}
 {\bf c}_0 \times {\bf c}_1 \neq {\bf 0} \label{eq:crossprod}
\end{equation}
then we know we have just formed a 2D wrapping. In higher dimensions, one might want to determine the span in some other way, such as forming a matrix out of all the wrapping vectors associated with a cluster and transforming it into row-echelon form.

Note that if we were to restrict ourselves only to planar lattices, the state of the wrapping would be a global property. For example, if we formed a wrapping with ${\bf c}=(1,0)$ after already having encountered a cluster with ${\bf c}=(0,1)$, then these clusters are necessarily attached and we have formed a 2D wrapping. However, in a non-planar lattice (including higher-dimensional lattices) it is possible to have disconnected non-colinear wrappings. We must therefore check the cross products of connection vectors every time we join clusters, with a 2D wrapping formed whenever (\ref{eq:crossprod}) is satisfied. Similarly, whether the lattice is planar or not we must be sure that the connection vector is inherited correctly when joining a 1D wrapping to a non-wrapping cluster.

\section{Convolution}
Finally, we need some means of numerically computing the convolution in Eq.~(\ref{eq:PA}). In the summation, the maximum $C$ is $L^2$ and $N$ is $\mathrm{O}(L^2)$, so this sum in principle contains $\mathrm{O}(L^4)$ terms for a two-dimensional lattice. In reality, not all these terms will be encountered by the sampling algorithm, as some states are either impossible or very unlikely. Rather, for each $n$ we will have data only between $C_{\mathrm{min}}(n)$ and $C_{\mathrm{max}}(n)$. We can consider either the theoretically possible $C_{\mathrm{min}}(n)$ and $C_{\mathrm{max}}(n)$ or the actually sampled $C_{\mathrm{min}}(n)$ and $C_{\mathrm{max}}(n)$, which will be a smaller range. Eliminating the impossible states from the summation divides the number of $(n,C)$ states only by a constant factor. Eliminating unsampled states (theoretically possible but very unlikely) gives a much larger improvement. To realize this improvement in memory and run time, sampled data is stored in a linked list instead of an array, so that we are only allocating memory for terms that have actually appeared. We then computationally find that the memory requirement is $\mathrm{O}(L^3)$. This memory requirement also follows theoretically from McDiarmid's bounded difference inequality, which applied here gives that the tail distribution on the probability of $C$ drops off like $e^{-O(C^2)}$ for $p\in(0,1)$ and $q=1$ \cite{McDiarmid}.

We also need to calculate binomial-type terms of the form
\begin{equation}
 B(N,n,C;p,q) \equiv {N \choose n} p^n (1-p)^{N-n}q^C . \label{eq:binom}
\end{equation}We can not evaluate (\ref{eq:binom}) directly, as individual terms may be very large. We follow NZ here by setting the largest term, at a point we denote $(n',C')$, to 1 and calculating all other terms relative to that with recursion. When $q=1$, the largest term occurs at $n'=N p$. However, for general $q$, we must take into account how $q^C$ skews the distribution of likely $n$. To find $(n',C')$, we can iterate through all sampled $(n,C)$ values, renormalize the maximum found so far to 1, and calculate further $B$ values by the recursive formulas that will be introduced below. In practice, we only need to iterate over one edge of the list, as for $q\le 1$ ($q\geq1$), the largest value of $B$ will come at the lowest (highest) sampled $C$ value for a given $n$, and thus this iteration will be $\mathrm{O}(L^2)$.

We assign this point $(n',C')$ to have a $B$ value of 1. We then use recursion to fill in the other terms:
\begin{equation}
B(N,n+1,C;p,q) = \frac{p(N-n)}{(1-p)(n+1)}B(N,n,C;p,q)
\end{equation}
\begin{equation}
B(N,n,C+1;p,q)=qB(N,n,C;p,q)
\end{equation}
In this way, we can compute all the relevant binomial coefficients subject to our arbitrary normalization. As long as we use these same coefficients to compute $Z$ in Eq.~(\ref{eq:Z}), Eq.~(\ref{eq:PA}) will be normalized correctly.

\section{Run-Time and Memory}
Our algorithm has three phases. The first is the sampling phase, which does not depend on $p$ or $q$. In this, we can run $t$ Newman-Ziff trial runs, each of which adds $\mathrm{O}(L^2)$ edges in averaged constant time each. Thus, this step takes approximately $\mathrm{O}(L^2t)$ time, though in practice it is more like $\mathrm{O}(L^{2.5}t)$ due to memory requirements. The linked list in which we store our data will have $\mathrm{O}(L^3)$ nodes, and thus we have an $\mathrm{O}(L^3)$ memory requirement, which the bounded difference inequality again shows will have a $t$-dependence of $\mathrm{O}(L^3\sqrt{\log(t)})$ (reflecting the fact that the size of the list must increase, although only slowly, as more rare configurations appear).

The next step is that, for a given $q$, we can do a pre-summation to reduce our run time and memory. Since the inner sum of Eq.~(\ref{eq:PA}) only depends on $q$, we can compute it before we know the $p$ value. In other words, we collapse the $(n,C)$ linked list to two arrays of length $N+1$, which store the sum and normalization for that $n$. To avoid overflow or underflow, we normalize to the highest sampled $C(n)$ (if $q\ge 1$, or lowest if $q\le 1$), which we also store in a third array for correct normalization in the post-summation step. This step requires $\mathrm{O}(L^3)$ time and reduces our memory from $\mathrm{O}(L^3)$ to $\mathrm{O}(L^2)$.

Then, we do a root-finding algorithm for this $q$ to find the critical $p$. The post-summation step for a given $(p,q)$ pair takes only $\mathrm{O}(L^2)$ time to sum over the arrays from the pre-summation step.

If the list of $q$ we are interested in is specified in advance, then it is possible to reduce the memory of the entire to $\mathrm{O}(L^2)$ by combining the sampling and pre-summation steps. This does require renormalizing as we sample based off the sampled $C(n)$, but takes approximately the same amount of time as the normal sampling algorithm.

\section{Results}
In Table \ref{tab:square}, we present results on the square lattice for various $q$, where the critical points are known to be \cite{Baxter1973}:
\begin{equation}
 p_c(q)=\frac{\sqrt{q}}{1+\sqrt{q}}.
\end{equation}
As mentioned earlier, if the critical point of a lattice is known exactly, the critical polynomial will find the exact solution for any size of basis. To illustrate this, calculations are presented for both $L=3$ using $10^{10}$ runs and $L=16$ using $10^9$. Although both are in agreement with the exact solution it is clear that for $L=16$ and $q>1$, accuracy suffers somewhat because of estimates that rely on some portion of $(n,C)$ space that is relatively difficult to resolve. We would expect this situation to improve eventually as $q$ becomes large because the probabilities become more weighted toward the least-occupied configurations. 

\begin{figure}
\begin{center}
 \includegraphics{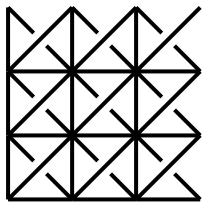}
 \caption{The $L=3$ basis for matching graph of the square lattice.}
 \label{fig:sq_match}
\end{center}
\end{figure}

As an example of an unsolved problem, we calculate critical points for the matching graph of the square lattice, the $L=3$ basis of which is shown in Figure \ref{fig:sq_match}. This is a non-planar lattice, with the diagonals in each face crossing but not directly connected to each other. Its site percolation threshold is $1-p_c(\mathrm{square\ site}) \approx 0.4073$, unsolved but known numerically to high precision \cite{Jacobsen15}. Thresholds are reported in Table \ref{tab:sq_match} for various $q$ including that for $q=1$, the bond percolation threshold, which has previously been found numerically by several authors \cite{FengDengBlote08, Ouyang2018} and very recently by Xu et al. \cite{Xu2021} using a Monte Carlo-critical polynomial technique. Their value is $0.250\,368\,40(4)$. Because none of these cases is exact, we expect the estimates to depend on $L$, and we used a few different sizes around $L=25$ just to confirm that the same predictions were made. These results were found with $10^8$ total runs, far short of what one might normally use but our goal is only to demonstrate that the algorithm works, not to present values of definitive accuracy. Note that one may instead want to do computations on smaller lattices and then extrapolate to infinite $L$, as this may be the best way to maximize the accuracy of the estimates. There is, however, not yet a good theory governing this extrapolation, and scaling exponents, which are known to be lattice-dependent \cite{Jacobsen15,ScullardJacobsen2020}, must be calculated empirically, so we do not explore this here.

\begin{table}
\begin{center}
 \begin{tabular}{l|l|l|l}
 $q$ & $L=3$ & $L=16$ & exact \\ \hline
 1.0 & 0.5000002(7) & 0.500000(2) & 1/2 \\
 1.5 & 0.5505103(6) & 0.550511(3) & 0.5505102572...\\
 2.5 & 0.6125740(5) & 0.61252(5) & 0.6125741133...\\
 3.5 & 0.6516683(4) & 0.6515(3) & 0.6516685226...\\
 9.5 & 0.7550342(2) & 0.7552(2) & 0.7550344704...\\
 10.0 & 0.7597467(2) & 0.7599(2) &0.7597469266...\\
 \end{tabular}
 \caption{Potts critical points, $p_c$, on the square lattice computed with the Monte Carlo algorithm for bases of size 3 and 16. These both agree with the known exact solution, as is standard for critical polynomials. These estimates are found by calculating averages of $P(\mathrm{2D}|n,C)$, $P(\mathrm{0D}|n,C)$, and $\Gamma_n(C)$, combining these with Eq. (\ref{eq:PA}) to get $P(\mathrm{2D})$ and $P(\mathrm{0D})$ and numerically finding the root of equation (\ref{eq:PB}). Results for all $q$ are found in a single simulation.}
 \label{tab:square}
\end{center}
\end{table}

\begin{table}
\begin{center}
 \begin{tabular}{l|l|l|l}
 $q$ & $p_c$ & $q$ & $p_c$ \\ \hline
 1.0 & 0.250368(2) & 7.0 & 0.4473(2)   \\
 1.5 & 0.28812(3) & 7.5 & 0.4546(2)   \\
 2.0 & 0.3164(4) &  8.0 & 0.46141(9)  \\
 2.5 & 0.3388(5) &  8.5 & 0.46778(9)  \\
 3.0 & 0.3577(5) &  9.0 & 0.47376(9)  \\ 
 3.5 & 0.3739(4) &  9.5 & 0.47940(9)  \\
 4.0 & 0.3881(3) &  10.0 & 0.48472(9)  \\
 4.5 & 0.4005(3) &  10.5 & 0.48976(9)  \\
 5.0 & 0.4117(3) &  11.0 & 0.49455(8)  \\
 5.5 & 0.4217(3) &  11.5 & 0.49910(8) \\
 6.0 & 0.4310(3) &  12.0 & 0.50345(8)  \\
 6.5 & 0.4395(2) &  &   \\
 \end{tabular}
 \caption{Critical points, $p_c$, for the random cluster on the square matching lattice.}
 \label{tab:sq_match}
\end{center}
\end{table}

\section{Discussion}
Although we have only studied the square and square matching lattices here, our algorithm works for all planar and non-planar periodic lattices and easily generalizes to higher dimensions. Its adaptability also gives it the potential to be used for related models, such as the Ashkin-Teller model \cite{AshkinTeller}, though for some generalizations such as multisite occupancy \cite{multisite}, the issue of a large number of disallowed edge configurations may hinder the algorithm's effectiveness. Our algorithm is also fast, as it can sample in $\mathrm{O}(L^{2.5}t)$ time for $t$ trials on an $L \times L$ lattice, and then find the critical point for each $q$ in $\mathrm{O}(L^3)$ time. The $\mathrm{O}(L^3\sqrt{\log{t}})$ memory requirement is steeper than some other algorithms, but this is only for the most general calculation; if a list of $q$ is specified in advance, the memory can be reduced to $\mathrm{O}(L^2)$. However, we also note that, because there is not much need to use lattices greater than $L \sim 30$ in 2D, memory might not be much of a consideration. A straightforward approach using arrays instead of a linked list carries an $\mathrm{O}(L^4)$ memory cost, and some might find this acceptable in exchange for a simpler implementation.

The main weakness of our algorithm is that, because we have separated the factor $q^C$ from the sampling and made $q$ a convolution parameter, some of the unsampled, ``unlikely'' configurations might prove to be important when their true weight is taken into account. Because our sampled region is of size $\mathrm{O}(L^3\sqrt{\log{t}})$ while the total possible region is of size $\mathrm{O}(L^4)$, this problem becomes more serious for large $L$. Increasing $t$ mitigates this problem, but only with the small effect of $\sqrt{\log{t}}$. How serious this is depends on the quantity being studied. For example, we have found that it is very difficult to accurately calculate the partition function, or free energy, for $q>1$ and large $L$, but the root of the critical polynomial still gives reasonable approximations to the critical point for $L=300$ (although such large calculations are hardly necessary). Thus, we mainly recommend the algorithm for finding critical points of the random cluster model, as it appears to be very useful for that purpose particularly if one wants to study multiple values of $q$ simultaneously.

The aforementioned problem is akin to the one seen in Hu's algorithm \cite{Hu1992}. There, sampling was done for $q=1$ and another value of $q$ could be obtained by a mapping which was subsequently \cite{Heringa1993} shown to behave poorly for large lattices. Although our method is very different, our sampling gives equal weight to all configurations with the same $n$, which is best suited to $q=1$. The ideal solution would be to find a sampling method that gives equal weight to all configurations with a given $(n,C)$, but where the weights of a state with $(n,C)$ and $(n,C+1)$ differ by a factor $1/q$. Unfortunately, it seems difficult to do this in an efficient manner. Wang, Kozan, and Swendsen \cite{Wang2002,Wang2003,WangConf} give a simple survival process that satisfies these criteria but they point out its extreme inefficiency when $q>1$. To address this, they devise a clever technique they call ``binary tree summation'', which is able to fill in a larger range of $(n,C)$ than our algorithm but  is also limited to small graphs and is otherwise not very general; for example, it would require significant modification to be able to measure wrappings to compute the critical polynomial. It remains to be seen whether a simple alternative exists that can fully extend the advantages of the Newman-Ziff algorithm to the random cluster model.

\section{Conclusion}
We have presented a Monte Carlo algorithm for the random cluster that, in the spirit of the Newman-Ziff algorithm for percolation, separates the edge and cluster weights $p$ and $q$ from the configuration sampling. The result is a method that can be used to calculate results corresponding to multiple values of real $q>0$ in a single simulation. Although it is generally limited to small graphs, using it to calculate the critical polynomial results in good estimates for the critical point, which for non-integer $q$ might be difficult to calculate with other methods.

\section{Acknowledgments}
Part of this research was performed while the authors were at the Institute for Pure and Applied Mathematics under funding from the National Science Foundation grant DMS-1925919. We thank IPAM and UCLA for hosting, and Susana Serna for her program leadership. We also thank Sina Zareian (Claremont Graduate University) for helpful conversations, advice, and resources. CRS thanks Robert Ziff for a stimulating and helpful discussion of this problem, and Martin Weigel for useful comments. THB is supported by NSF Graduate Research Fellowship Grant Nos. DGE-1745016 and DGE-2140739. This work was partially performed under the auspices of the U.S. Department of Energy at the Lawrence Livermore National Laboratory under Contract No.\ DE-AC52-07NA27344 and was supported by the LLNL-LDRD Program under Project No.\ 19-DR-013.

\section*{References}
\bibliographystyle{iopart-num}
\bibliography{scullard.bib}
\end{document}